\documentclass[article]{elsarticle}

\usepackage{hyperref}
\usepackage[english]{babel}
\usepackage{layouts}
\usepackage{amsmath}
\usepackage{xcolor}

\journal{Journal of \LaTeX\ Templates}

\begin{document}

\begin{frontmatter}

\title{Numerical simulation of the EM forced flow during Sn-Pb alloy directional horizontal solidification}
\tnotetext[mytitlenote]{Fully documented templates are available in the elsarticle package on \href{http://www.ctan.org/tex-archive/macros/latex/contrib/elsarticle}{CTAN}.}

\author{E. Shvydkiy\fnref{myfootnote}\cortext[mycorrespondingauthor]}
\address{Ural Federal University, Mira 19, 620002, Yekaterinburg, Russia}
\ead{e.l.shvydky@urfu.ru}

\author{I. Smolyanov}
\address{Ural Federal University, Mira 19, 620002, Yekaterinburg, Russia}

\author{E. Baake}
\address{Leibniz University of Hannover, Wilhelm-Busch-Str. 4, 30167, Hannover, Germany}



\begin{abstract}
This article presents a study of the influence of electromagnetically (EM) forced convection on segregation formation.
As a reference case the solidification of Sn-5 wt pct Pb alloy from the Hebditch and Hunt experiment is taken.
Applied forcing of convective flow is determined by the dimensionless electromagnetic forcing parameter $F$. 
The study was carried out in the range $-3.5\times 10^6<F<3.5\times 10^6$.
Velocity profiles and flow patterns in the liquid phase are obtained for different applied EM forcing conditions.
As a result of parametric analysis, the dependence of the Reynolds number on the electromagnetic forcing parameter was obtained.
Electromagnetic forces can significantly affect the flow in the liquid bulk, increasing and slowing down the velocity, as well as changing the circulating flow direction.
Solute concentration distribution analysis has shown that EM-forced convection does not affect global solute segregation formation. 
For the two cases, segregation channels were obtained.
However, the analysis of the global segregation index showed that an increase in the Reynolds number provokes a slight decrease in this parameter.
The main mechanism of segregation formation is transport of solute rich liquid into the mushy zone.
In considered numerical experiment configuration, the penetration into the mushy zone is not large, and even a change in the direction of convection in the liquid bulk does not significantly affect the mushy zone flow.
The calculations were made by means of open source code in OpenFOAM and Elmer.

\end{abstract}

\begin{keyword}
EM-forced flow\sep columnar solidification\sep Hebditch and Hunt experiment \sep macrosegregations \sep convection \sep numerical simulation

\end{keyword}

\end{frontmatter}





\section{Introduction}

Phase transition processes occur in a number of industrial applications such as casting, crystal growth, welding, and others.
Particularly in the production of metallic materials, the solidification step is responsible for the formation of macrostructure and defects such as heterogeneities of alloy composition (macro- and microsegregations).
Macrostructure formation can be controlled by grain refiner or increasing cooling rate, but in some cases their use may be limited or impossible.

The solidification process is carried out by undercooling the liquid metal below the melting point.
In this case, temperature and concentration gradients occur.
These gradients create buoyancy forces and in many casting technologies these forces induce convective flow in the liquid bulk.
These convection flows play an important role in solidification processes and have a significant and multifaceted effect on the structure and segregation formation during solidification \cite{flemings1974solidification,kurz1984fundamentals,dantzig2016solidification,Nikrityuk2011}.
For example, convection may be responsible for the shape of the transitional mushy zone, its microstructure, and, finally, for macrosegregation. 
However, the relationship between flow and solidification is still one of the major unknown \cite{DavidsonARFM1999}.

Typically, convection in liquid metal can be caused by buoyancy forces (natural convection)  due to the difference in fluid density, which in turn arises from temperature or concentration gradients.
The intensity of such natural convective flows is dimensionlessly evaluated.
In contrast to the Rayleigh number widely used in the fluid mechanics community, in works dedicated to convection during solidification, as the defining non-dimensional parameter the Grashof number is used. 
The thermal (index $_T$) or concentrational (index $_C$) Grashof number determines the ratio of the buoyancy to viscous force acting on a fluid defined as follows:

\begin{equation}
	Gr_T= \frac{g \beta_T \Delta T H^3}{\nu^2},~~~~~Gr_C= \frac{g \beta_C \Delta C H^3}{\nu^2},
	\label{Gr_T}
\end{equation}
where $g$ is the gravity vector, $\beta$ is the expansion coefficient, T is the temperature, $H$ is the characteristic size, $\nu$ is the kinematic viscosity and $C$ is the solute concentration.

The buoyancy number that characterizes the importance of solutal buoyancy forces compared with thermal ones. The buoyancy number $N_b$ is expressed as follows:

\begin{equation}
N_b= \frac{ \beta_C \Delta C}{\beta_T \Delta T}
\label{N_b}
\end{equation}

Contactless free-convective flows in the liquid phase can be controlled using electromagnetic forces \cite{Eckert2013}.
The use of alternating electromagnetic fields makes it possible to generate induced currents and, consequently, electromagnetic forces directly in the liquid metal. These body forces can effectively control convective flows during solidification.
This method has proven itself in metallurgical applications, due to the contactless nature of this method, and the ability to create magnetic fields of various configurations and strengths.

To bring the value of electromagnetic forces to a dimensionless form and compare them with buoyancy forces, the dimensionless electromagnetic forcing parameter is introduced \cite {Dubke1988_Part_I,Dubke1988_Part_II,ShvydkiyMMTB2021}.
\begin{equation}
F = \frac{\left< F_{x} \right> H^3}{\rho \nu^2},
\label{F}
\end{equation}
where $\left< F_{x} \right>$ is volume averaged horizontal component of electromagnetic body force and $\rho$ is the density.

A number of authors \cite{GRANTS2004630,Noeppel2010,Hachani2015,Avnaim2018a} have proposed using parameter $N=\frac{F}{Gr} =\frac{\left< F_{x} \right> }{ \rho g\beta\Delta T }$ to estimate the ratio of two convective forces.
The value of this parameter determines the dominant flow generation mechanism. At  $N<1$ corresponds to the case of thermal convection, and at $N>1$ -- to the case of electromagnetic stirring.
However, the issue of the influence of this ratio on the solidification of pure metal was only partially considered in \cite{Avnaim2018a}.
A more detailed study of the influence of this parameter on the crystallization of pure metal, and, in particular, of a two-component alloy, has not yet been carried out.

Previous studies on the effect of electromagnetic stirring on the solidification process have been based mainly on an empirical approach. 
They consisted mainly in determining the influence of the intensity and configuration of the electromagnetic field on the structure of cast ingots.
In this case, the experiments formulation was based on setups close to industrial casting facilities.
However, with this approach, firstly, the dimensionless criteria for the  magnetic field effect on the structure are not defined, and secondly, the experiment design, which is close to real installations, introduces additional effects.
This situation prompted researchers to look for more simplified research formulations.

One such widespread research formulation is based on an experimental setup for the study of macrosegregations formation during horizontal solidification in a rectangular cell, first proposed by Hebditch and Hunt \cite{Hebditch1974}.
Recently, Hachani et al. \cite{Hachani2015} created a setup for studying EM-forced convection during solidification similar to the Hebditch and Hunt experiment.
The known installation was supplemented with a traveling magnetic field inductor, which generates electromagnetic forces in the solidifying alloy.
Thus, it is possible to study the influence of natural convection and electromagnetically forced flow on the solidification process.
It should be noted that this research formulation and the chosen type of experimental setup were recognized as successful and later identical experimental setups were created in different laboratories \cite{Avnaim2018,Avnaim2018a,LOSEV2019125249}.
Based on the experimental data obtained in this experiment, a bench of numerical models  are successfully validated and tested (see for example \cite{BoussaaIJHMT2016,ZhengIJHMT2018,ChenMet2019,WangIJHMT2020,KumarIJHMT2021,AbdelhakemJHT2022,KhelfiJCG2022}).

In this experiment on directional horizontal solidification, the electromagnetic influence on the grain structure (transition from columnar to equiaxed), as well as macro and mesosegregations results were obtained.
It has been confirmed that the use of electromagnetic forcing makes it possible to influence the mechanism of formation of segregations and a fine equiaxed structure, as well as the ability to avoid segregation channels and reduce macrosegregation \cite{Hachani2015}.
However, before obtaining exhaustive data describing the mechanism of the influence of flows caused by external electromagnetic force action is still far away.
This requires numerous studies describing the influence of many factors on the conditions for the formation of micro and macrostructures for various metals and alloys.

Since solidification is a complex multiscale process, the second aspect considered in this article is numerical simulation.
In parallel with the development of experimental studies, there was a rapid development of new numerical methods.
And in the field of materials processing, numerical simulation tools are increasingly being used.
The prediction of temperature fields and liquid metal velocity fields using numerical models is currently quite reliable. 
Modeling individual aspects of metallurgical processes has become a daily practice for engineers working in industry.

However, the multiphase, multiscale, and multiphysical nature of the solidification process often limits the prediction of even the most advanced modeling tools.
Current knowledge and modeling tools are still limited and considering new more complex models, expansion of existing numerical codes is badly needed at present \cite{LudwigMMTA2015} and will increase knowledge about important process details and vice versa \cite{LudwigMMTB2014}.
Simulation of magnetohydrodynamic (MHD) flow is currently the only way to evaluate the phenomena that occur during electromagnetic stirring.
At this stage, most of the research is carried out experimentally, but more scientific effort is still needed to build a model capable of relating MHD phenomena and the formation of segregation during solidification \cite{LudwigJOM2016}.
In this work, an open source model will be presented which is capable of calculating columnar solidification coupled with electromagnetic stirring.

Our hypothesis is that by controlling the convection parameters in the liquid phase of a solidifying ingot using an electromagnetic field, it is possible to influence the formation of the micro and macrostructure of the ingot.
This, in turn, will provide the necessary knowledge about the desired solidification conditions to achieve the specified parameters of the cast metal ingots.

\section{Methods}

To create a model capable of calculating electromagnetic stirring in coupling with the solidification of a binary alloy, we refer to the literature. It was decided to build a new model based on coupling two existing computational codes:
\begin{itemize}
	\item Electromagnetic stirring (\textit{Elmer + OpenFOAM})\footnote{\url{https://github.com/jvencels/EOF-Library}} \cite{Vencels2017}:
	\item Columnar solidification (\textit{OpenFOAM})\footnote{\url{https://github.com/OpenFOAM/OpenFOAM-Solidification}} \cite{Coleman2020MST,ColemanThesi2020}
\end{itemize}

The first part of the model is based on a combination of Elmer and OpenFOAM softwares by means of EOF Library \cite{Vencels2019}. In this part, the time-harmonic electromagnetic (EM) field generated by the inductor and the force field in the volume of liquid metal are calculated by the finite element method. This EM force field is interpolated and sent from the electromagnetic part to the fluid dynamic one. Magnetic field advection by melt velocity \cite{Azulay2018PF} is not taken into account due to a low magnetic Reynolds number.

\begin{equation}
\mathbf {F_{em}} =0.5 \Re\{\underline{\mathbf{J}} \times  \underline{\mathbf{B}} \}, 
\label{F_{em}}
\end{equation}

where $\mathbf{J}$ is the current density and $\mathbf{B}$ is the magnetic flux density.

However, this resulting electromagnetic force is not the only source body force term in the momentum equation. Moreover, during the solidification process, the presence of thermal and concentration gradients in the bulk of the liquid is expected. Therefore, the fluid is also affected by the buoyancy forces $ \mathbf {F_ {source}} = \mathbf {F_ {em}} + \mathbf {F_b} $.
Where the buoyancy force $\mathbf{F_b}$ is calculated by the Boussinesq approximation and includes not only thermal expansion but also the influence of solute concentration:
\begin{equation}
\mathbf{F_b}=\mathbf{g} \rho = \mathbf{g} \rho_0(1 - \beta_T(T - T_{0}) - \beta_C(C - C_{0})),
\label{F_b}
\end{equation}

The solidification of a binary alloy is modeled using the two-phase (incompressible) averaged solidification method \cite{WuMetals2019} .
The model uses a solver based on the continuum mixture theory taken from \cite{BENNON19872161}.
This approach is reliable in determining the final pattern of segregations, and in particular, segregation channels.
This model is implemented in OpenFOAM based on the finite volume method \cite{Coleman2020MST}.

\begin{figure}[h]
	\begin{minipage}[h]{0.53\linewidth}
		\centering
		\includegraphics[width=1.0\linewidth]{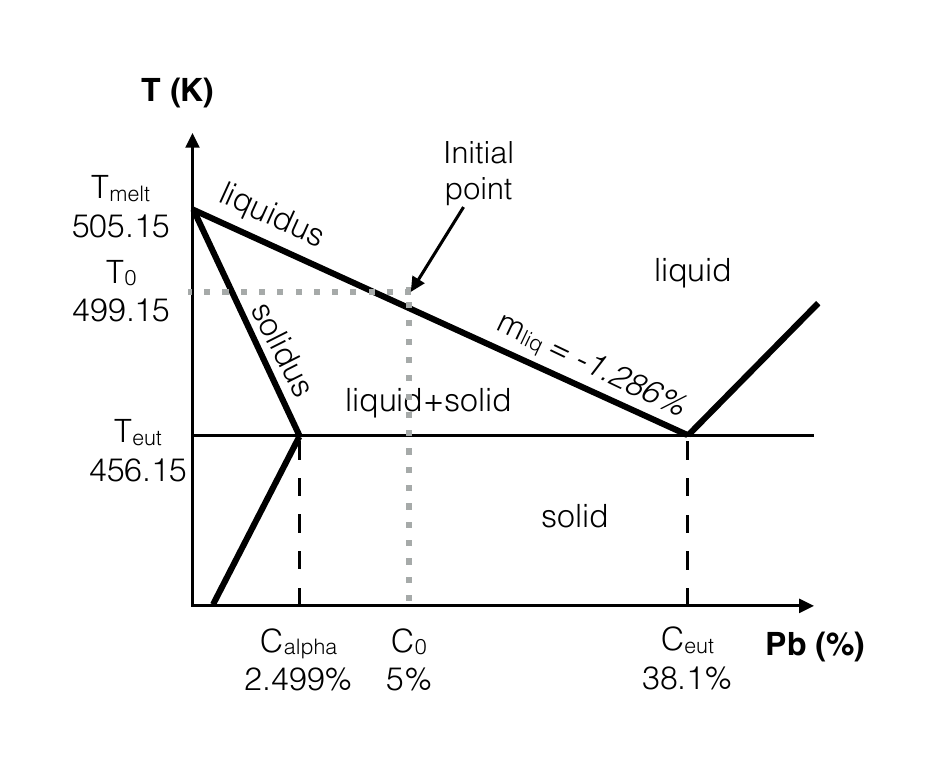}\\  
	\end{minipage}
	\hfill
	\begin{minipage}[h]{0.47\linewidth}
		\centering
		\includegraphics[width=1.0\linewidth]{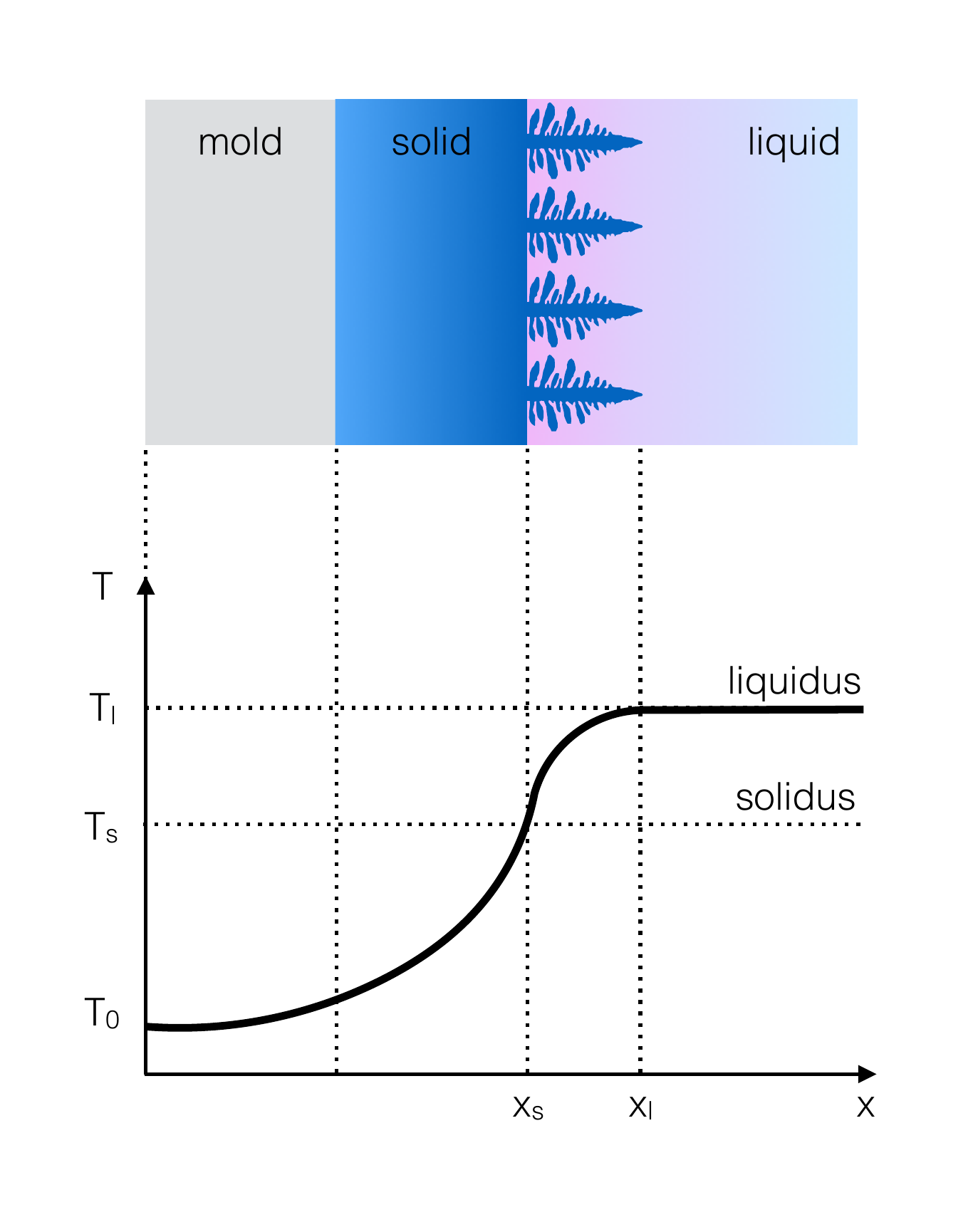}\\  
	\end{minipage}
	\hfill
	\caption{Principal scheme of a linearized phase diagram of a binary Sn-Pb alloy} \label{PhaseDiagram} 
\end{figure}

One of the describing phase transformations for alloys basics is the phase diagram. In this case, the phase transition from liquid to solid and vice versa depends not only on temperature, but also on the solute concentration. The considered two phase PbSn alloy system will be treated by the phase diagram, which is shown in Figure \ref {PhaseDiagram}. This phase diagram is linearized and the properties are presented in Table \ref {Properties}.

Another feature that we will mention is the transitional solid liquid zone or so-called mushy zone. In the process of binary alloy solidification, the growth of dendrites likely occurs \cite {flemings1974solidification}. The volume-averaged method of alloy solidification modeling \cite {WuMetals2019} is widespread to solve such phenomena on the macro level. The model chosen is the non-dendritic columnar solidification model (two phase model). In this model, the solid phase is considered as a columnar solid phase. 
A transition zone between solid and liquid phases, also called a mushy zone, is treated as a porous medium, which makes it possible to simulate the movement of liquid metal in this zone.
The flow in this porous mushy zone is taken into account by adding a Darcy term \cite{Kumar_2013}  in the momentum equation with a permeability $K$ given by a Kozeny-Carman law based on the secondary dendrite arm spacing (DAS) $\lambda$. 
\begin{equation}
K= \frac{\lambda^2}{180} \frac{g_l^{3}}{(1-g_l)^{2}} 
\end{equation}
where $g_l$ is the volume fraction of liquid phase.

To summarize, the momentum equation of the liquid velocity during solidification of a binary alloy with electromagnetic stirring will take the following form:

\begin{equation}
\frac{\partial \mathbf{U}}{\partial t} +\nabla \cdot g_l \mathbf{U}+ (\mathbf{U} \cdot \nabla) \mathbf{U} - \nu \nabla^2 \mathbf{U}= \frac{\mathbf{F_{source}}}{\rho_0}-\frac{\mu g_l}{K \rho_0}\mathbf{U},
\end{equation}
\label{momentumEq}
where $\mathbf{U}$ is the velocity, $t$ is the time, $\nu$ is the kinematic viscosity.





The problem formulation is two-dimensional. Taking into account recommendations (for example from \cite{LI2012407a}) the numerical grid with $250 \times 150$ elements was used.
The computation time is 1400 s with the solver time step is set 0.005 s.

\subsubsection{Electromagnetic field equations}
Electromagnetic forces were calculated using Elmer open-source software. The general equations to calculate electromagnetic field variable can be presented by
electric scalar $\varphi$ and vector magnetic scalars $\mathbf{A}$ in transient form
\begin{equation}
\Delta \mathbf{A}- \mu \sigma \dfrac{\partial \mathbf{A}}{\partial t} -\mu \sigma \nabla \varphi+\mu \sigma (\mathbf{U}\times \nabla \times \mathbf{A})=-\mu \mathbf{J},	
\label{eq:Aphi}
\end{equation}
\begin{equation}
\mathbf{B} = \nabla \times \mathbf{A},
\label{eq:B}
\end{equation}
\begin{equation}
\mathbf{J} = \sigma \left(-\nabla \varphi + \dfrac{\partial \mathbf{A}}{\partial t} \right),
\label{eq:J}
\end{equation}
and then to rewrite in harmonic form using Laplace transformation 
\begin{equation}
\Delta \mathbf{\underline{A}}-j\omega \mu \sigma \mathbf{\underline{A}}-\mu \sigma \nabla \underline{\varphi}+\mu \sigma (\mathbf{U}\times \nabla \times \mathbf{\underline{A}})=-\mu \mathbf{\underline{J}},
\label{eq:Aphi}
\end{equation}
\begin{equation}
\mathbf{\underline{B}} = \nabla \times \mathbf{\underline{A}},
\label{eq:B}
\end{equation}
\begin{equation}
\mathbf{\underline{J}} = \sigma (-\nabla \underline{\varphi} + j \omega \mathbf{\underline{A}}).
\label{eq:J}
\end{equation}
Here, $\omega$ is the angular velocity of magnetic field expressed as $ 2 \pi f $ using frequency of magnetic field $f $, 
$\mu$ is the absolute magnetic permeability, 
$ \sigma $ is the conductivity, 
$\mathbf{J} $ is the current density, 
$ \mathbf{B} $ is the magnetic flux density and
$j$ is the imaginary unit. 
Underlining of a number depicts belonging the number signifies that the number belongs to a complex number area.
This approach makes it possible to significantly reduce computing resources, because it is not necessary to calculate magnetic field parameters for each time step. 
Magnetic field advection by melt velocity is not taken into account due to a low magnetic Reynolds number.

As the thermal boundary conditions for the left side wall, heat extraction is set ($h=300~Wm^{-2}K^{-1}$, $T=289.15~K$). All remaining walls are adiabatic (thermal insulation) and the velocity at all walls is zero ($u=0$)
Initially, the whole domain is liquid ($g_l=1$, $u=0$) slightly above the liquids line ($T=499.15~K$,  $C_0=0.05$).
All thermophysical parameters are assumed to be constants, that is, they do not depend on temperature or phase.

\section{Verification and Validation}

The main problem of the chosen solidification modeling approach is the uncertainty of the numerical parameters, in particular, the permeability of the two-phase mushy zone.
In order to get the reliable result, it is necessary to test the model on well-known benchmarks.

\subsection{Verification}

To verify the model for calculating columnar solidification, we refer to the well-known SMACS numerical benchmark for 2D solidification of binary alloys \cite {BELLET20092013}.
This case is a rectangular symmetric mold ($0.06\times  0.1$ m) with Pb-18~wt~\%~Sn alloy.
Both side walls are cooled, while the remaining ones have thermal insulation boundary conditions.

\begin{figure}[h!]
	\begin{minipage}[h]{0.5\linewidth}
		\centering
		\includegraphics[width=1.0\linewidth]{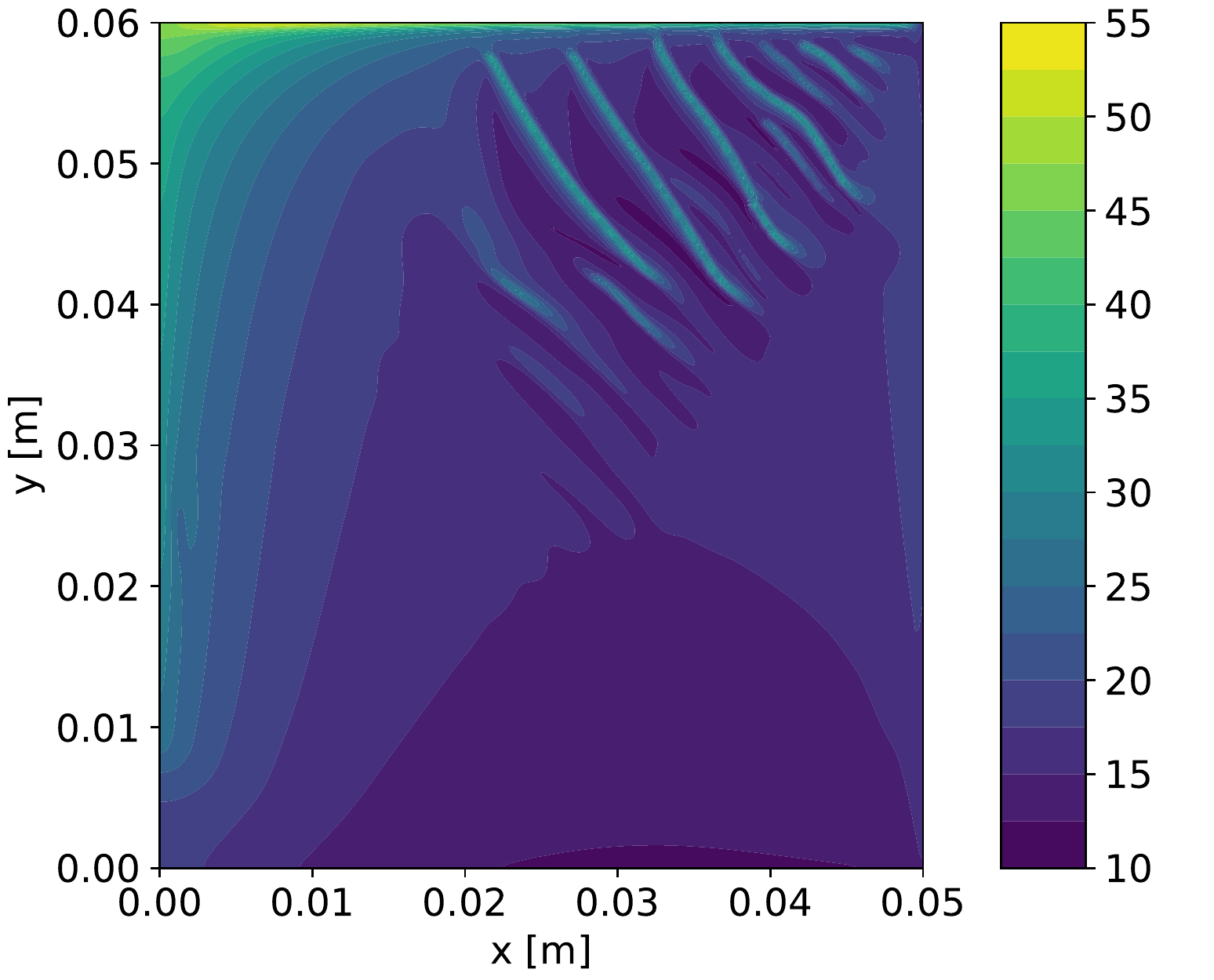}\\  a
	\end{minipage}
	\hfill
	\begin{minipage}[h]{0.5\linewidth}
		\centering
		\includegraphics[width=1.0\linewidth]{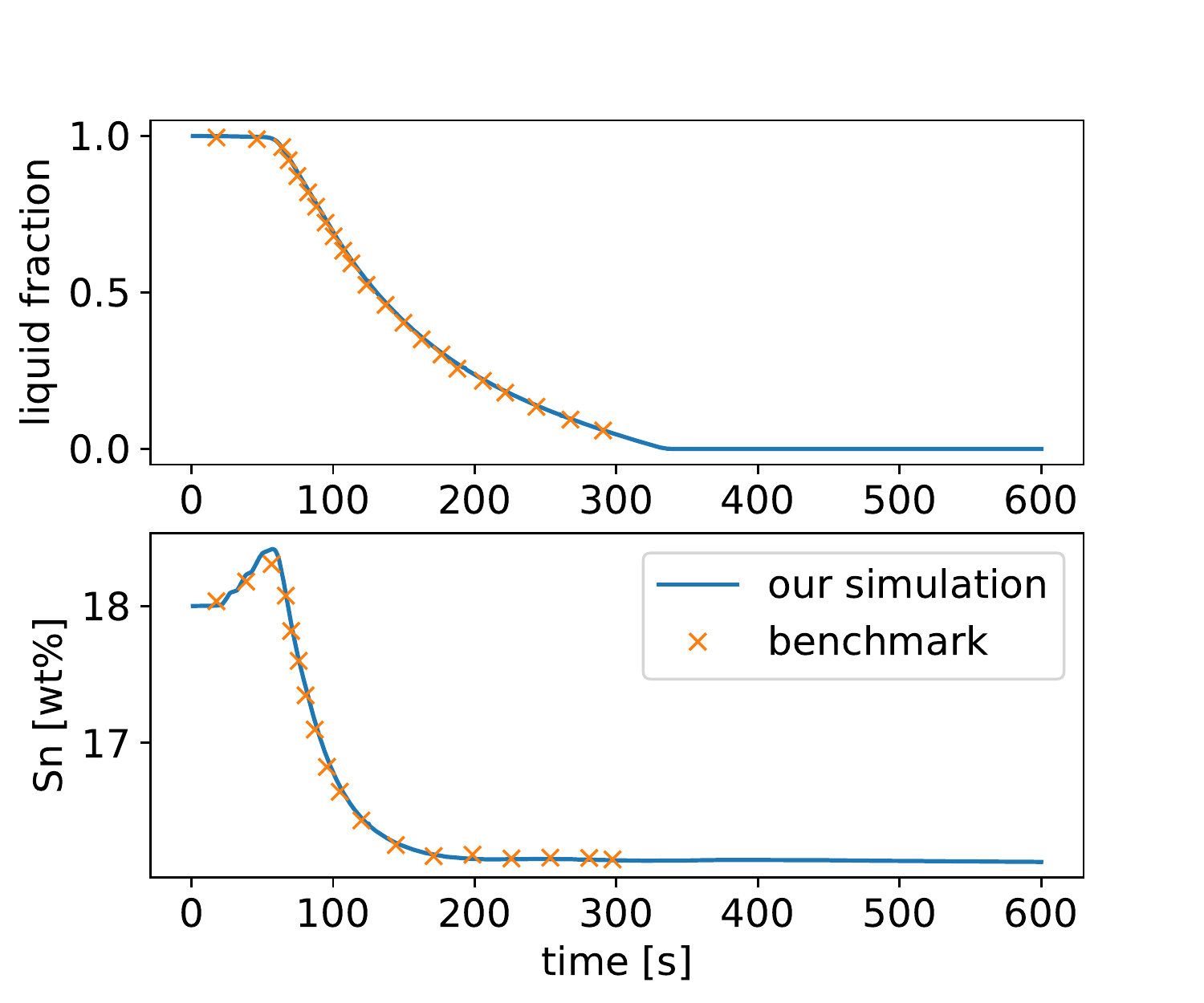}\\  b
	\end{minipage}
	\hfill
	\caption{(a) Final Sn composition predictions for the SMACS case and (b) comparison of the solute mass fraction and liquid mass fraction at the centre of the domain to numerical benchmark case Ref. \cite{Combeau_2012}.} \label{Verification} 
\end{figure}

The figure \ref {Verification} shows the results of this benchmark case simulation.
In these circumstances, solute-driven convection producing a counter clockwise flow will dominate.
The Sn concentration field at the 600 s of solidification process is shown in Figure \ref{Verification}(a).
One central macro-segregation and several segregation channels can be observed in the ingot. This concentration distribution corresponds to the results obtained by various computational codes presented in the work \cite {Combeau_2012}.
A comparison of solute mass fraction and liquid mass fraction histories at the domain midpoint are compared to \cite {Combeau_2012} in Figure \ref {Verification} (b), with a very good agreement.

\subsection{Validation by means of the Hebditch and Hunt experiment}

\begin{table}[h]
	\caption{Thermophysical data used in the Calculations}
	\label{Properties}	
	\label{tab_1}
	\begin{center}
		\begin{tabular}{lcll}
			
			\hline
			\bf	Parameter & \bf Symbol & Value& Units \\
			\hline
			
			Melting temperature& $T_{melt} $ &505.15 &  $ K$ \\ 
			Eutectic temperature& $T_{eut} $ &   456.15& $ K$ \\ 
			Eutectic concentration& $C_{eut} $ & 0.381   &\\ 
			Alpha phase point & $C_{\alpha} $ &0.025&    \\ 
			Density	&$\rho_0$ & 7000&kg/m$^3$   \\ 
			Specific heat capacity &$C_p$ &260& $ J kg^{-1} K^{-1}$ \\ 
			Thermal conductivity	&$\kappa$ & 55& $ W/(m\cdot K)$   \\ 
			Dynamic viscosity	&$\mu$ & 1e$^{-3}$ &$ Pa\cdot s$  \\ 
			Mass diffusion coefficient	&$D$ & 1e$^{-8}$& $ m^2\cdot s^{-1}$  \\ 
			Dendrite arm spacing &$\lambda$ & 0.2e$^{-3}$& $m$ \\ 		
			Thermal expansion coefficient	&$\beta_T$ & 6e$^{-5}$ &$ K^{-1}$  \\ 		
			Solutal expansion coefficient	&$\beta_C$ & -5.3e$^{-2}$ & \\ 
			Reference temperature	&$T_{0}$ & 499.15& $K$  \\ 						
			Reference concentration	&$C_{0}$ & 0.05& \\ 		
			Latent heat	&$L_{a}$ & 6.1e$^{4}$ &$K kg^{-1}$  \\ 								
			Electrical conductivity&$\sigma$ &2e$^4$  &S/m   \\ \hline										
		\end{tabular}
	\end{center}
	\label{properties}
\end{table}

\begin{figure}[h!]
	\center
	{\includegraphics[width=1.0\linewidth]{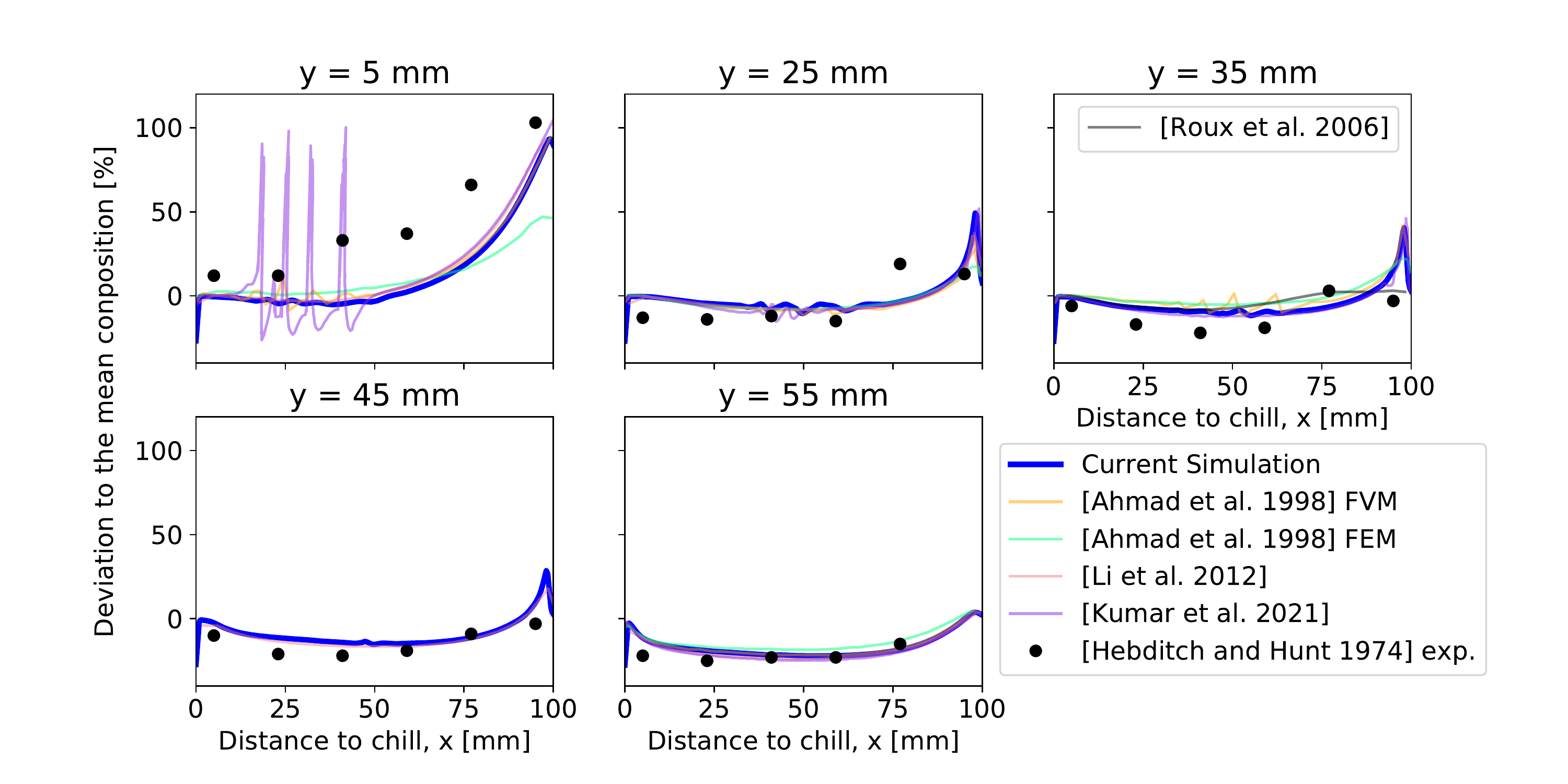}}
	\caption{Deviation to the mean composition from the Hebditch and Hunt experiment. References here: Hebditch and Hunt 1974 \cite{Hebditch1974} Ahmad et al. 1998 \cite{AhmadMMTA1998}, Roux et al. 2006 \cite{RouxIJHMT2006}, Li et al. 2012 \cite{LI2012407a}, Kumar et al. 2021 \cite{KumarIJHMT2021}. }
	\label{HH_Validation}
\end{figure}

The classic experiment made by Hebditch and Hunt \cite{Hebditch1974} was chosen to validate the model.
In this experiment Sn-5 wt pct Pb alloy was solidified directionally in a thin section mold (0.06 m high, 0.1 m long and 0.013 m thick). Heat was extracted from only one side wall.
An estimate was made of the spatial distribution of the tin concentration at several different heights of the ingot.
The result is deviations of the mean concentration for different heights which are shown in the figure \ref{HH_Validation}.
Moreover, the literature analysis showed that several computational codes were also validated on Hebditch and Hunt data. We also processed the data from these works \cite{AhmadMMTA1998,RouxIJHMT2006,LI2012407a,KumarIJHMT2021} and plotted the results on this graph.

The results shown by our model have certain deviations with the experimental data in the case of $y=5$ mm.
But in the same case, the numerical results of other authors also have a similar discrepancy.
For other cases, the results of our simulation are in good agreement with the experimental data.

\begin{figure}[h!]
	\center
	{\includegraphics[width=1.0\linewidth]{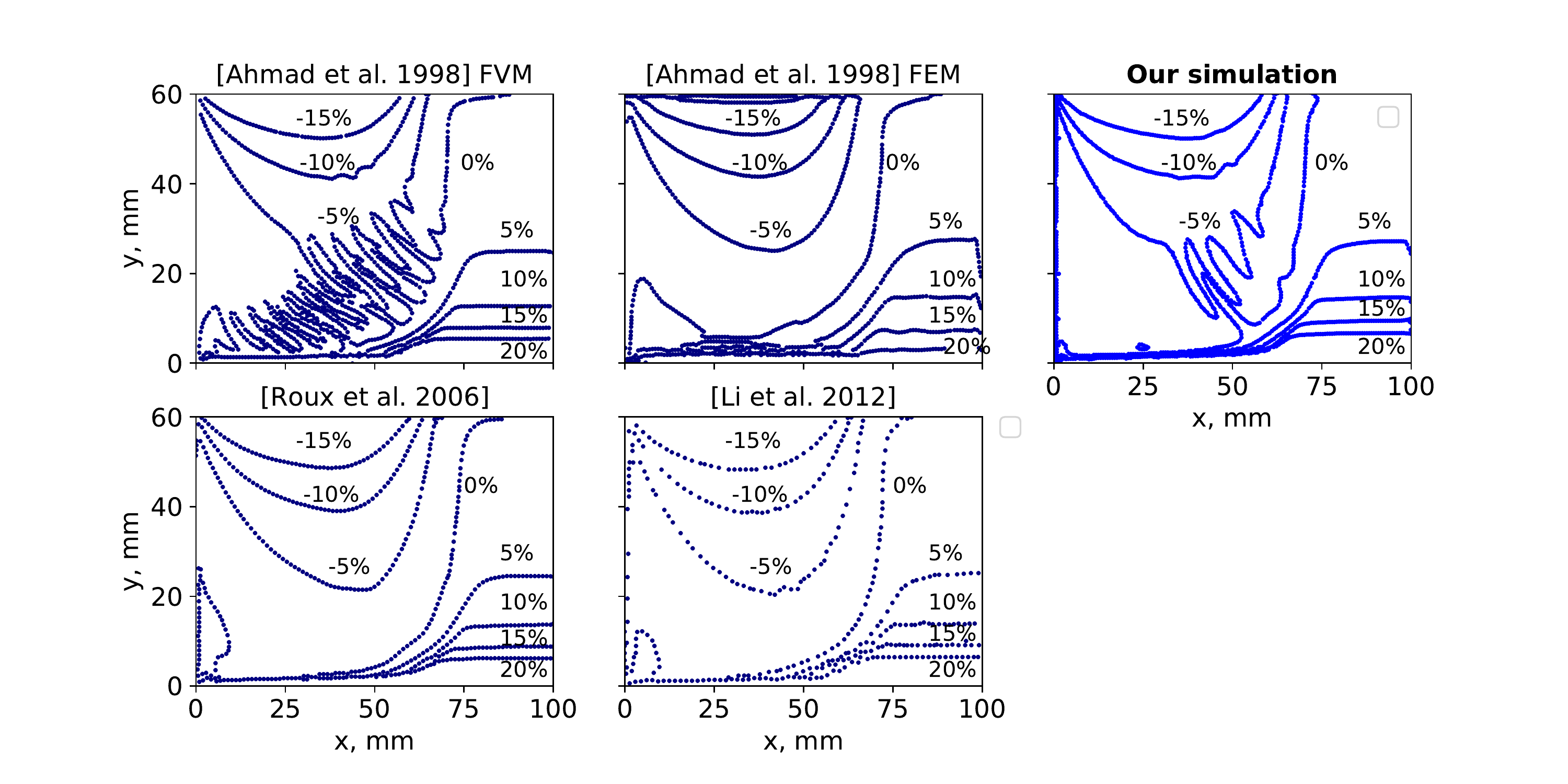}}
	\caption{Predicted segregation patterns in the Sn-5 wt.\% Pb benchmark at 400 s. References here: Ahmad et al. 1998 \cite{AhmadMMTA1998}, Roux et al. 2006 \cite{RouxIJHMT2006}, Li et al. 2012 \cite{LI2012407a}.}
	\label{HH_Verification}
\end{figure}

An additional assessment of the concentration distribution was also made by comparison with the results of other authors.
Figure \ref{HH_Verification} shows predicted segregation patterns for benchmark at 400 s.
The model presented in this work agrees quite well with the known results of other authors.
The only notable difference is the prediction of a trend towards segregation channels in the central region of the ingot.
In the case of [Ahmad et al. 1998] FVM (finite volume method) there is a strong ribbing of the contours 0\% and -5\%, while in other examples this is not observed.
Our model also predicts a channeling trend, however not as pronounced.

\section{Electromagnetic stirring}

\begin{figure}[h!]
	\center
	{\includegraphics[width=0.6\linewidth]{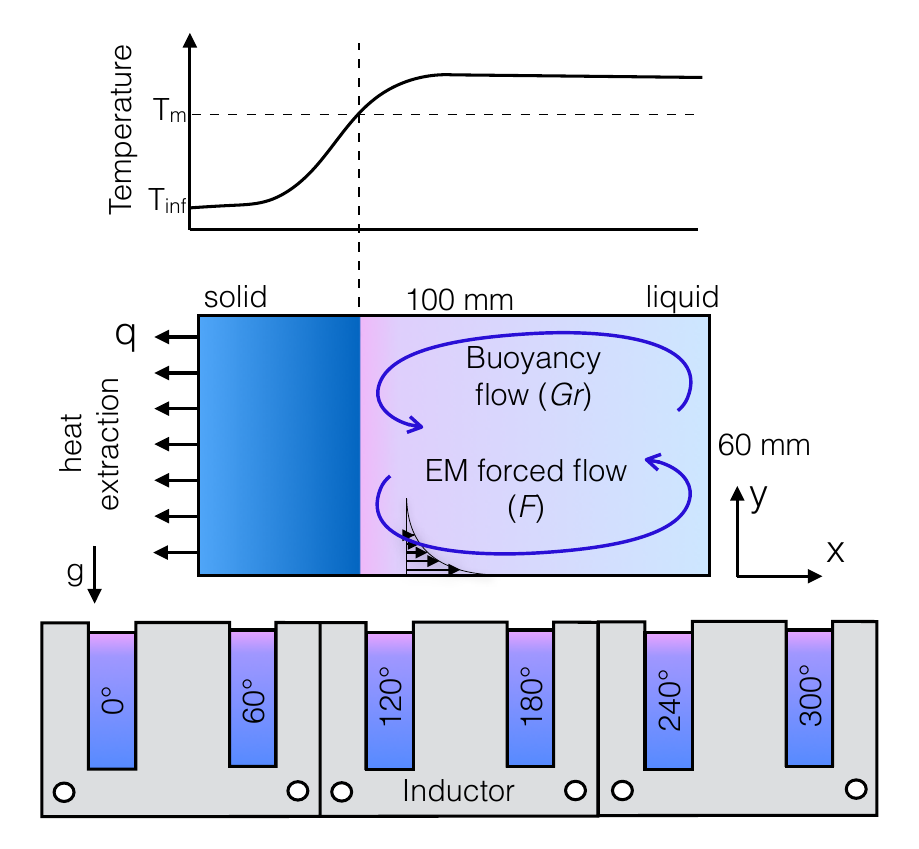}}
	\caption{Numerical experiment sketch.}
	\label{expSetup}
\end{figure}

In order to investigate the influence of EM-driven flow on the solidification process in the considered classical experiment Hebdich and Hunt the magnetic field inductor is implemented.
This inductor is a stator of a linear induction machine, which produces a travelling magnetic field (TMF).
This field penetrates into the lower area of the solidified alloy and induces the eddy-currents.
Interaction of these currents with TMF produces electromagnetic forces. 
These forces are the source of EM-forced flow into the liquid bulk (see Eq. \ref{F_{em}}).
A sketch of the described experimental setup is shown on Fig. \ref{expSetup}.
The detailed inductor parameters cal be found at \cite{Shvydkiy_metals}.
Similar experimental setups to study the impact of EM-driven flow on the solidification process by means of TMF are used in several laboratories \cite {Hachani2015, Avnaim2018a, LOSEV2019125249}.

\section{Results}
\subsection{EM forces}

\begin{figure}[h!]
	\begin{minipage}[h]{0.5\linewidth}
		\centering
		\includegraphics[width=1.0\linewidth]{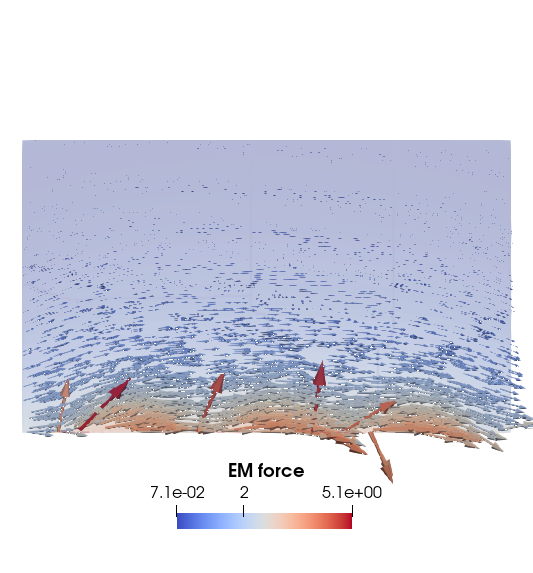}\\  a
	\end{minipage}
	\hfill
	\begin{minipage}[h]{0.5\linewidth}
		\centering
		\includegraphics[width=1.0\linewidth]{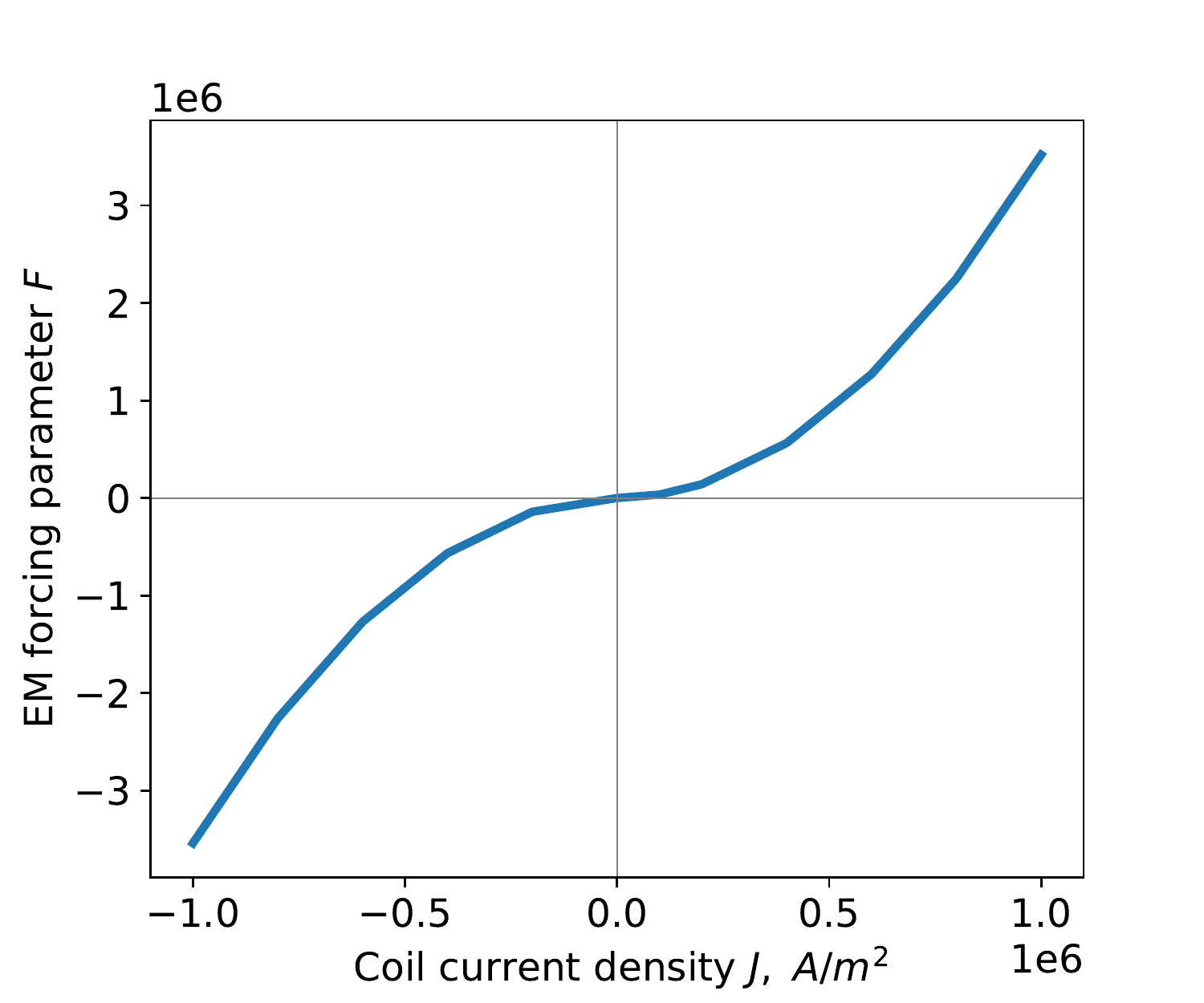}\\  b
	\end{minipage}
	\hfill
	\caption{EM forces in the calculated solidifying alloy domain (a) and resulting EM forcing parameter over the coil current density (b).} \label{EMforces} 
\end{figure}

The result of the electromagnetic part calculation, which is exported for further simulation in OpenFOAM, is the field of electromagnetic forces. Such induced forces are the product of magnetic flux density and eddy currents in the solidified alloy. The figure \ref{EMforces}a shows the distribution of the vector force field for the case under consideration.
Due to the presence of the skin-effect, these forces are concentrated at the lower boundary, closest to the inductor.
The direction of the forces is to the right, which corresponds to the direction of the traveling magnetic field generated by the inductor.
Obtained forces configuration should provide counterclockwise liquid metal bulk flow.

Moreover, a parametric study was carried out during which the value and direction (sign) of the current density in the inductor coils were varied. Figure \ref {EMforces}b shows the dependency of the EM forcing parameter for different coil current densities. This EM forcing parameter is calculated according to equation \ref {F}. 
The resulting dependence is close to a cubic function and with increasing current density parameter $ F $ will increase nonlinearly. At the negative coil current density the electromagnetic forcing  $ F $ parameter will also change its direction to opposite.

\subsection{Convection into the liquid phase}

\begin{figure}[t]
	\begin{minipage}[h]{0.5\linewidth}
		\centering
		\includegraphics[width=1.0\linewidth]{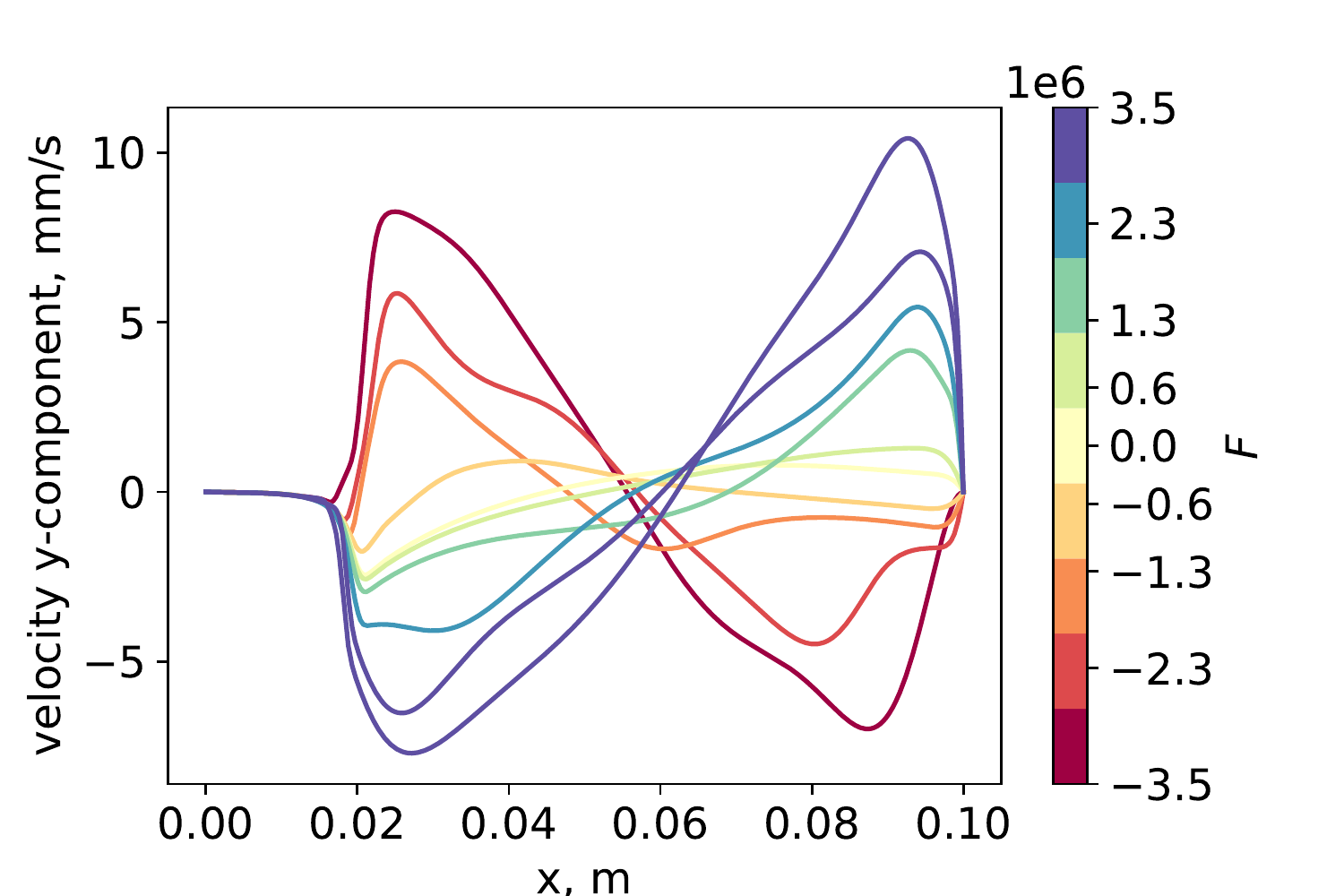}\\  a
	\end{minipage}
	\hfill
	\begin{minipage}[h]{0.5\linewidth}
		\centering
		\includegraphics[width=1.0\linewidth]{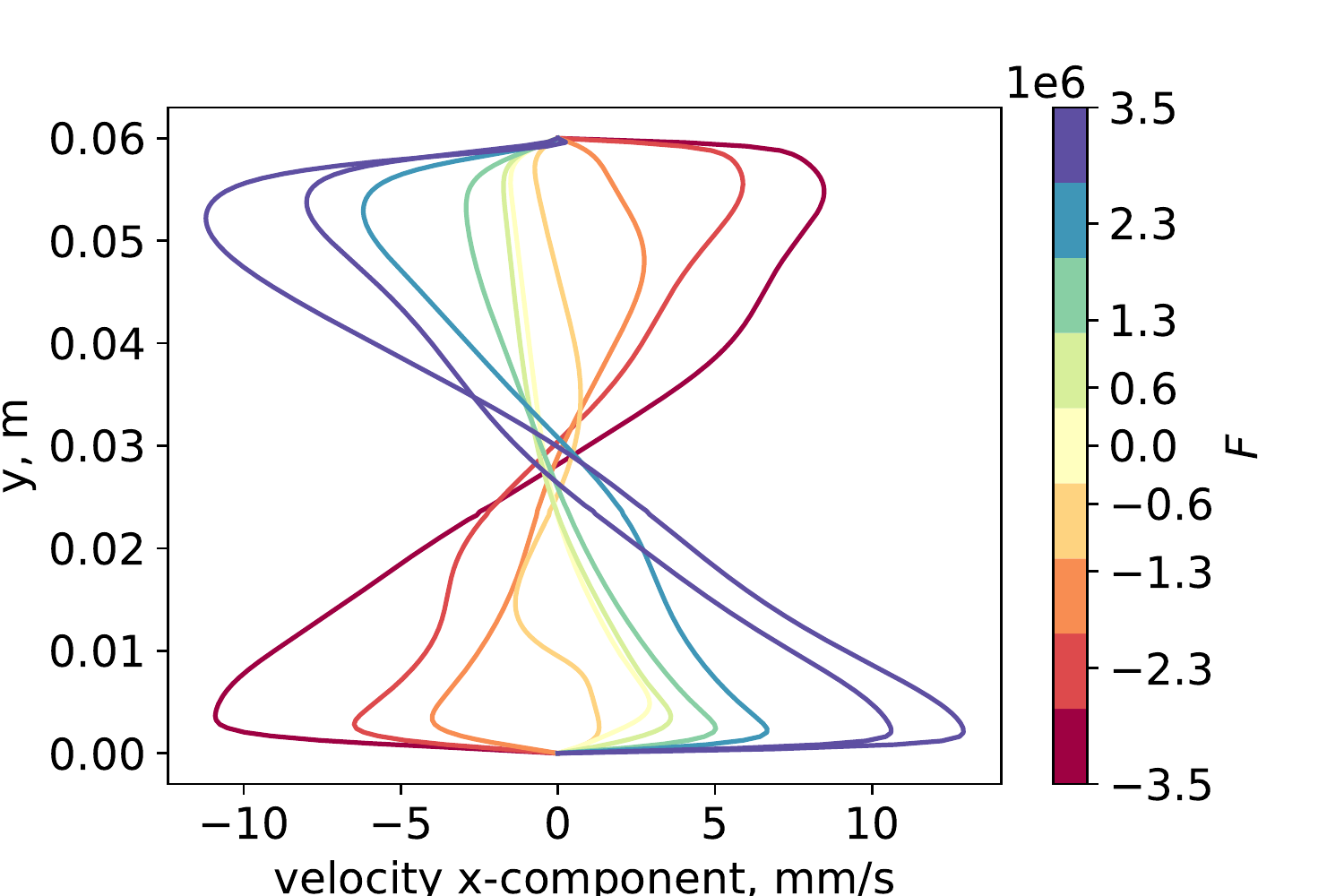}\\  b
	\end{minipage}
	\hfill
	\caption{Vertical (a) and horizontal (b) velocity profiles over the mold middle lines for different EM forcing parameters} \label{v_yAndv_x} 
\end{figure}

The electromagnetic forces obtained as a result of the electromagnetic calculation will be one of the body force terms in the momentum equation (\ref {momentumEq}). 
Thus, their magnitude and direction will affect the flow in the liquid phase of the solidifying ingot.
Fig. \ref {v_yAndv_x} shows the velocity profiles at 50 sec of the solidification process.
The vertical component is shown in the Fig. \ref {v_yAndv_x}a and the horizontal one in the Fig. \ref {v_yAndv_x}b over the horizontal and vertical midlines of mold respectively.
First of all, at the value of $ F = 0 $ there is only natural convection and a counter clock wise (ccw) flow with a maximal velocity less than 5 mm/s is observed.
The initial direction of the electromagnetic forces coincides with the buoyancy forces, therefore, as the parameter $ F $ increases, the vortex velocity increases at a constant direction.
The graph shows velocity profiles for different values of $ F $ and at $ F = 3.5\cdot10^6 $ the maximum velocity will increase several times and exceed the value of 10 mm/s.

At the negative value of $F$, electromagnetic forces will be directed against the buoyancy forces.
As can be seen from the graph, this will lead to a decrease in the natural convection velocity and the appearance of additional vortices (at $F=-6\cdot10^5$).
With a further negative increase in the value of $ F $, the flow turns and circulates in the opposite direction.
At the $F<-1.3\cdot10^6$ EM driven flow fully predominates over the natural convection and the max velocity reaches -10 mm/s. 

\begin{figure}[h]
	\center
	{\includegraphics[width=1.0\linewidth]{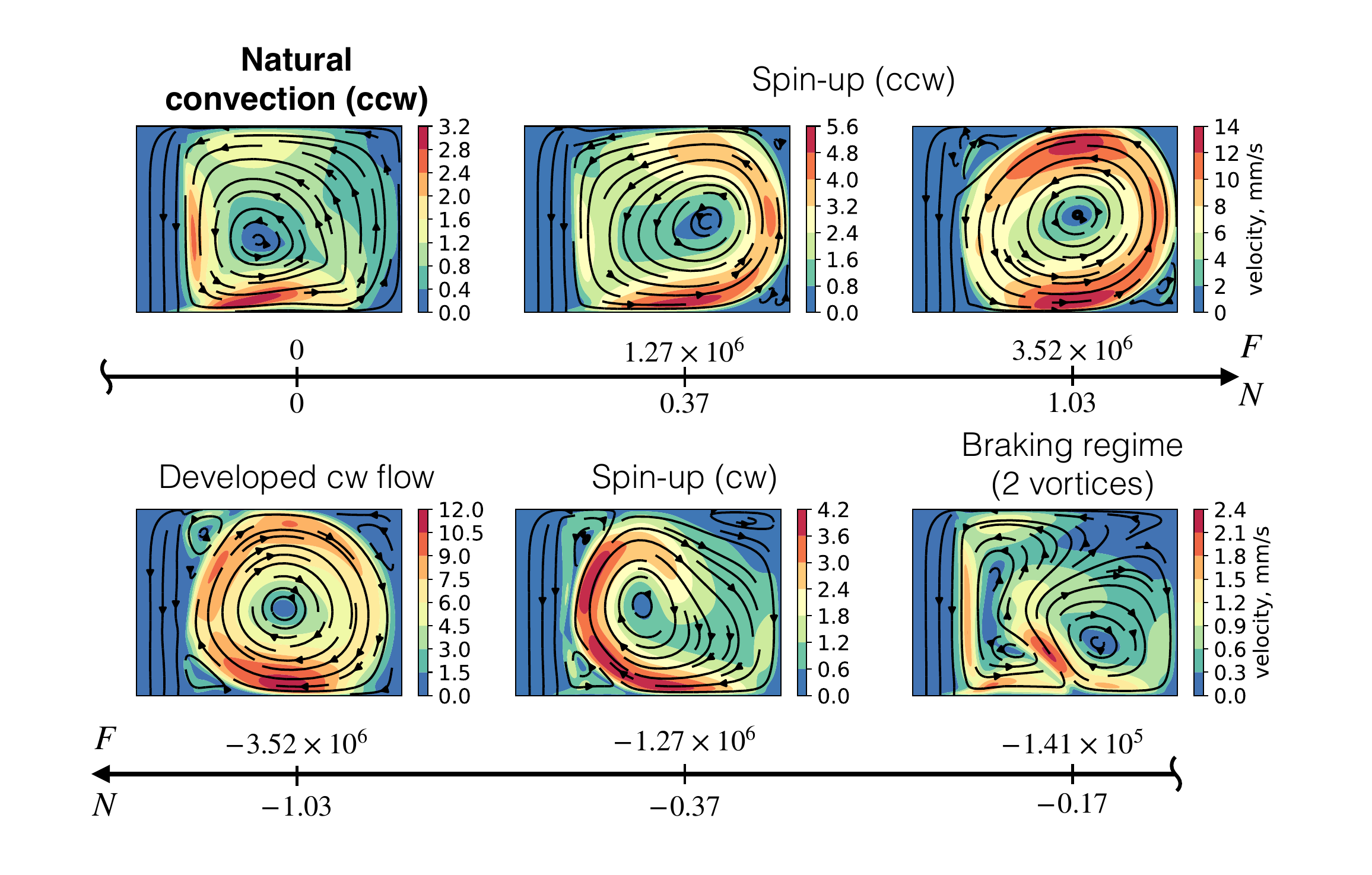}}
	\caption{Flow patterns for differing EM forcing conditions after 50 s of solidification.}
	\label{FlowPatterns}
\end{figure}

Additionally to the graphs, the 2D plots of flow patterns are shown in Fig.~\ref {FlowPatterns}.
For the case of natural convection ($F=0$), one circulated eddy is observed as well as in the results shown in Fig. \ref {v_yAndv_x}.
As $F$ increases, the velocity of this vortex grows, reaching a value of 14 mm/s.
However, this vortex does not significally deform the liquid solid interface.
With negative $F$ values the EM-forced flow is directed opposite to natural convection and at $F=-1.41\cdot10^5 $ braking mode can be observed. 
Namely, there are simultaneously two vortices (natural and EM forced) with low velocities $\sim2$ mm/s.
With an even greater increase in electromagnetic forces ($ F = -1.27 \cdot 10^6 $) EM driven flow starts to predominate over the thermal convection and one clockwise (cw) flow is developed.
Lastly for maximum negative forcing ($F=-3.52\cdot 10^6$) a cw flow with a slightly lower velocity in comparison to $F=3.52\cdot 10^6$ case is obtained.
It should be noted that the shape of the vortices tends to be round, this was noted in the works \cite {BOTTON201353, HAMZAOUI2019167} and is a consequence of the 2D formulation.
This form of flows is not entirely reliable, and in this work there is a particular goal to evaluate the interaction of the flow on the segregations formation, and in order to obtain more accurate flow structure, it is necessary to carry out full 3D simulations.

The Reynolds number is one of the integral characteristics of convection, describing the intensity of convective flow.
There are several ways to estimate the Reynolds number of circulating flow in a closed container. As a characteristic vortex size, $H$, $L$, or even $H^2/L$  can be used.
Since our cell is longer in x direction and moreover thermal and concentration gradients are applied along x axis we will evaluate Reynolds number in the following way \cite{TeimurazovPRF2017}:
\begin{equation}
Re=\frac{\langle |v_x|\rangle_V H}{\nu}
\label{Re_x_avg}
\end{equation}
where $u_x$ is the velocity horizontal component and $\langle  \rangle_V$ denotes averaging over volume. 

\begin{figure}[h]
	\center
	{\includegraphics[width=8cm]{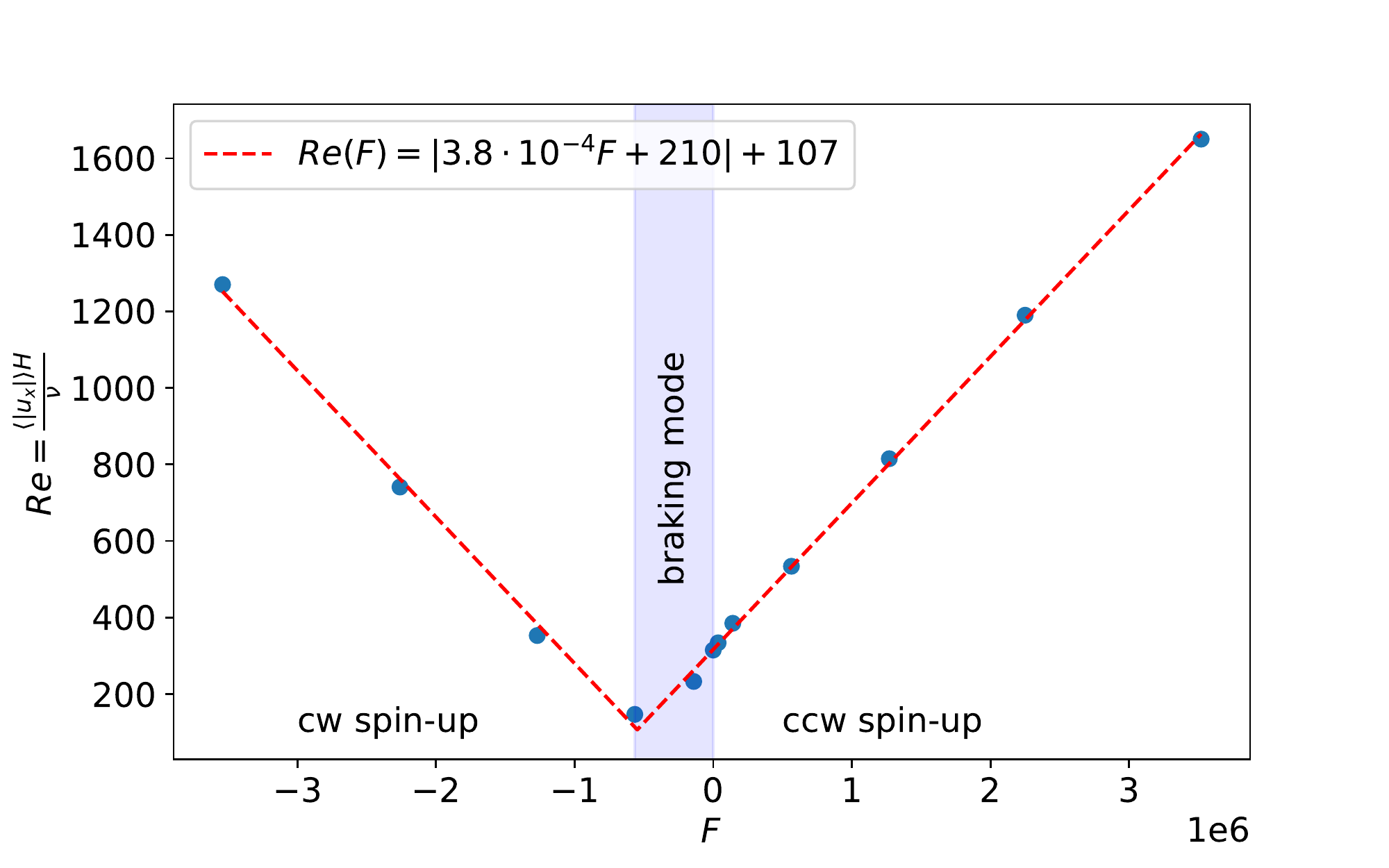}}
	\caption{Reynolds number over the EM forcing parameter $F$ at 50 s of solidification time.}
	\label{Re(F)aprx}
\end{figure}

The figure \ref{Re(F)aprx} shows the dependence of the Reynolds number on the EM forcing parameter.
It can be seen from the graph that as $F$ increases, the Reynolds number grows linearly, reaching a value of 1600.
This range corresponds to $F \geq 0 $ and is marked on the graph as ccw spin-up regime.
If a negative $F$ is applied, then with a small value of electromagnetic forces, the Reynolds number will be reduced to $\sim100$.
This braking mode is in the range $-5.36 \cdot 10^{5}\leq F \leq 0 $.
And with a further increase in negative $F$, we get cw spin-up regime in the range of $F \leq -5.36 \cdot 10^{5}$.
The obtained data can be approximated by the function $Re(F)=|3.8 \cdot 10^{-4}F+210|+107$.

\subsection{Solute concentration}

\begin{figure}[h]
	\center
	{\includegraphics[width=1.0\linewidth]{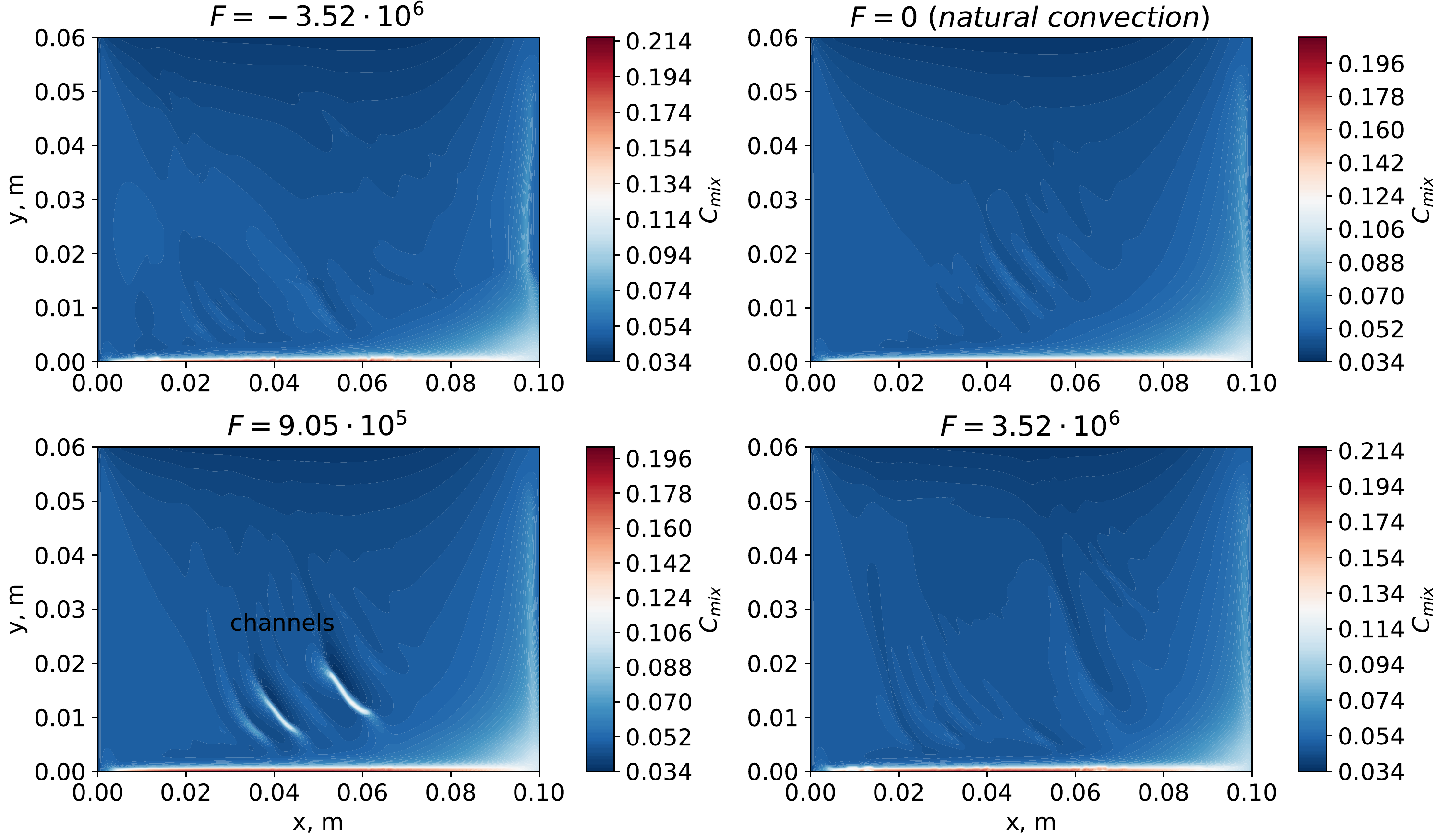}}
	\caption{Final concentration map for different EM forcing conditions.}
	\label{ConcentrationContourPlot}
\end{figure}

The final solute concentration maps for different forcing cases are shown in figure \ref{ConcentrationContourPlot}.
First of all, it should be noted that a significant change in the concentration distribution is not observed in any case.
The lead rich metal has a higher density and tends to concentrate at the lower boundary.
A vertical macrosegregation is formed at the right wall.
It can be noted only for the case $F=-3.52\cdot10^6$ that this vertical segregation is shifted to the left, which can be caused by means of cw flow.

Differences in these small spatial concentration variations are the only consequence of EM forced flow.
Neither an increase in velocity or even a change in the direction of flow into the liquid bulk led to a significant redistribution of concentration field.

In all cases, in the middle of the ingot, there is a tendency to form channels, and in the case of $F=9.05\cdot10^5$, two channels are completely formed.
Segregation channels in this study are considered to be an area with a deviation of the mean concentration $> 50\%$.
In two cases, segregation channels were obtained.
For the case $F=9.05\cdot 10^5$, the channel length was 6.53 mm, and its slope relative to the horizon was $140^{\circ}$. At $F=1.27\cdot 10^6$ the channel length is 1.6 mm and slope angle is $125^{\circ}$.
Although such a phenomenon is obtained in this study, the effect of EM-forced convection on channel formation mechanism\cite{Sample1984TheMO,LI2012419} should be further investigated.


However, this approach to analysis is not complete. For a more accurate assessment, it is proposed to use the Global Segregation Index (GSI) \cite{KumarIJHMT2021} which is defined as follows:

\begin{equation}
GSI=\frac{1}{C_0}\Bigg[\frac{1}{V_{domain}} \iiint_{V_{domain}}^{} (C-C_0)^2 dV\Bigg]^{\frac{1}{2}}
\end{equation}

\begin{figure}[h]
	\center
	{\includegraphics[width=7cm]{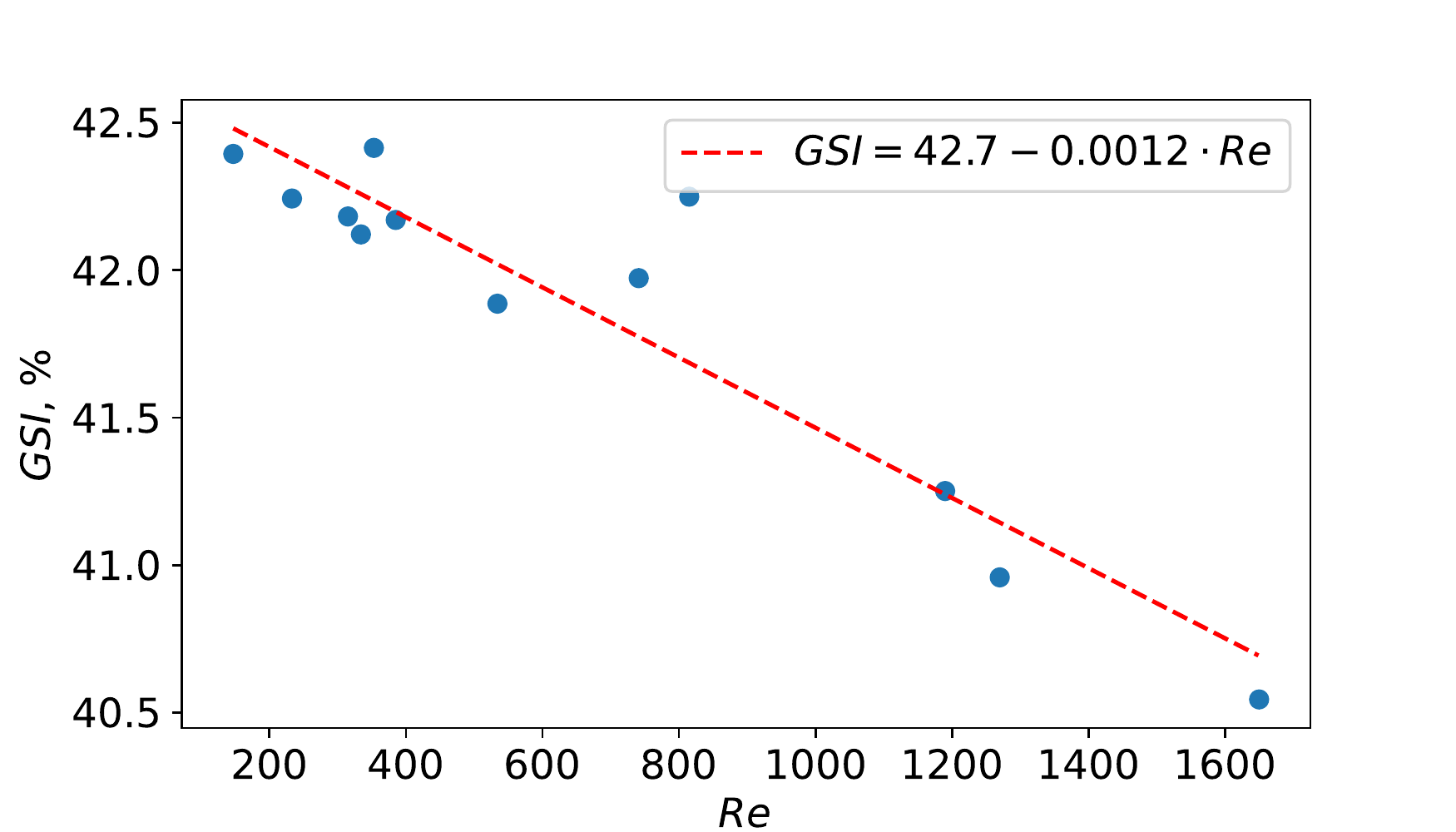}}
	\caption{Global Segregation Index ($GSI$) as a function of Reynolds number $Re$.}
	\label{GSI(Re)apr}
\end{figure}

Fig. \ref{GSI(Re)apr} shows the Global Segregation Index over the Reynolds number scatter plot. Obtained data is varied from 40.5\% to 42.5\% and approximated by the linear function $GSI(Re)=42.7-1.2Re\times 10^{-3}$. This means that at high Re number values segregation index tends to decrease. After the substitution the obtained dependence of global segregation index from the EM forcing parameter can be written as $GSI(F)=42.6-4.6 \times |F|\times 10^{-7}$.

\subsection{Flow into the mushy zone}

Previous results have shown that changing the flow velocity in the liquid bulk does not make a significant contribution to the spatial mixture concentration distribution.
Even reversing the flow direction only makes minor changes.
Whereas convection in the liquid phase should play a significant role in the formation of the macrostructure.
The influence of convective flow on the formation of macrosegregations is explained mainly by the washing effect near the liquid-solid interface.
In other words, this phenomenon is characterized by the presence of a flow in the mushy zone that transfers the rejected solute rich liquid to the liquid bulk.

In order to analyze the effect of electromagnetic stirring on the formation of macrosegregations, it is necessary to evaluate the relationship between the flow velocity in the liquid bulk and into the mushy zone.

We can calculate the vertical component of velocity into the mushy zone as follows






\begin{equation}
U_{mush~i}=\langle U_y \rangle_{V_{i}}
\end{equation}

where $i=0.8,~0.9,~0.98$ and $V_i \in 0.5 (1-i)<g_l<0.5 (1+i)$.

To evaluate the mushy zone volume we take three representative cases: 0.8, 0.9, 0.98. In this way it will be possible to estimate how deep into the mushy zone the main liquid bulk flow can penetrate.

\begin{figure}[h]
	\center
	{\includegraphics[width=9cm]{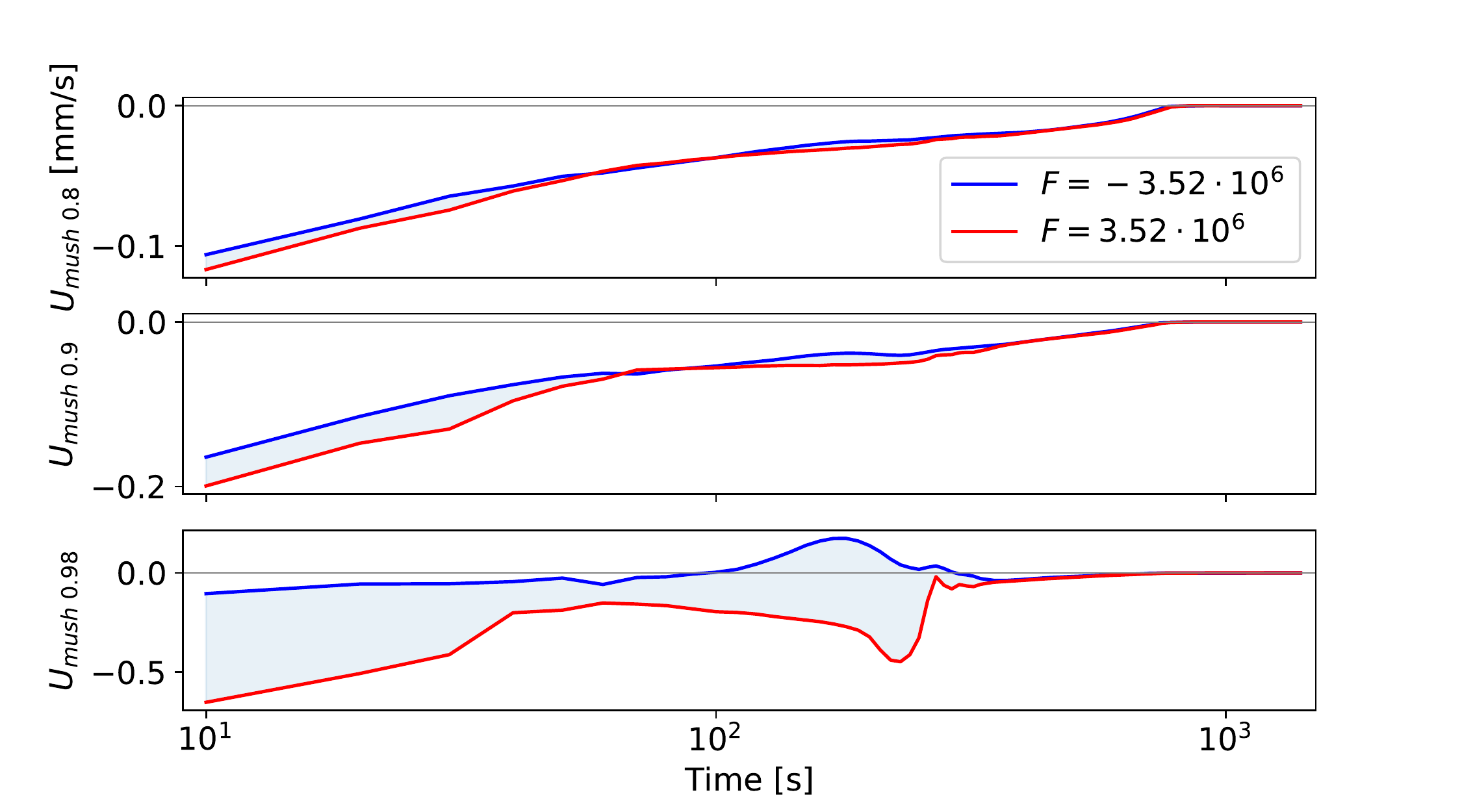}}
	\caption{Integral velocity dynamics into mushy zone.}
	\label{V_mush}
\end{figure}

For comparative analysis, we took two cases with maximum $F$ directed in opposite directions.
The Fig. \ref{V_mush} shows the vertical velocity dynamics in the mushy zone for these two cases.
At the same time, in the liquid bulk, developed flows of oppositely directed ccw and cw are observed (see Fig. \ref{v_yAndv_x} and \ref{FlowPatterns}).
For the $U_{mush~0.98}$ case, the greatest difference in the curves is observed.
The minimum velocity at 10 s is -0.7 mm/s for $F=3.52\cdot 10^6$.
While at $F=-3.52\cdot 10^6$ the initial velocity is -0.1 mm/s.
Moreover, in the range of 100-300 seconds, there is a positive direction of the speed into the mushy zone.

The graphs for $U_{mush~0.9}$ and $U_{mush~0.8}$ are qualitatively similar.
The initial velocity value for $U_{mush~0.9}=-0.2$ mm/s and for $U_{mush~0.8}=-0.1$ mm/s respectively.
Then the flow rate in the mushy zone gradually decays.
The differences between $F=3.52\cdot 10^6$ and $F=-3.52\cdot 10^6$ do not exceed 20\%.
For all considered cases at 350 s, the velocities for positive and negative $F$ become equal, decay smoothly, and already after 800 s are equal to zero.

Based on the displayed mushy zone velocity dynamics, we can conclude that the change of convection in the liquid bulk to the opposite allows a significant effect only on the shallow layers of the mushy zone ($1>g_l>0.95$).
In the larger volume of the mushy zone ($0>g_l>0.95$), EM-driven flow does not penetrate and does not make significant changes.

Since the rejected transport into the mushy zone is a crucial factor of segregation formation,
further research on the effect of forced convection on segregation formation must aim on flow control into the mushy zone.
Moreover, in the model used, the permeability of the porous medium in the mushy zone depends on such numerical parameters as DAS taken from the literature. 
And more detailed studies of the effect of permeability on the interaction of the two velocities considered are needed \cite{ZhangMTC2020}.

\section{Conclusion}
In this work the parametric numerical study of the electromagnetic forcing influence on binary alloy solidification is presented.
The main conclusions can be formulated as follows:

\begin{itemize}
\item Open source code for calculation of binary columnar solidification under the electromagnetic impact is presented.
This model is verified and validated on existing numerical and experimental data.
We are open to contributions to improve this model (for example equiaxed grain formation (CET) interface) here \footnote{\url{https://github.com/shvydkiy/EM-Stirring-Convection-Solidification/tree/main/EM\%2BSnPbSolidification}}.

\item Velocity profiles and flow patterns in the liquid phase are obtained for different applied EM forcing conditions.
Electromagnetic forces allow spin-up, break-down or reverse the natural convection flow.
As a qualitative result, the dependence of the Reynolds number on electromagnetic forcing parameter is obtained. 

\item The final segregation patterns for different EM forced conditions are obtained.
Simulation results did not show evidence of dramatic changes in concentration field distribution. 
The fact that neither an increase in the velocity nor even a change in the direction of flow into the liquid bulk led to a significant redistribution of concentrations can be explained in the following ways.
Firstly, the developed flow is not present for a sufficiently long time compared to the full period of solidification.
Secondly, numerical parameters such as DAS are used to calculate the permeability of the mushy zone and it is possible that the flow from the liquid bulk to the mushy zone is not so intense.
	
\item Segregation channels are obtained for the two cases ccw EM driven flow.
	
\item Global segregation index analysis has shown that GSI depends linearly on Re number and constituently on EM forcing parameter.
	
\item Velocity analysis in the mushy zone showed the reason why electromagnetic stirring did not significantly change the concentration distribution.
It was found that the velocity in the liquid phase does not directly affect the mushy zone flow $U_{liquid~bulk}\neq U_{mush}$.

\end{itemize}

\section*{References}

\bibliography{elsarticle-IJHMT}

\newpage
\appendix
Appendix 1: Dynamics parameters.
\begin{figure}[h!]
	\center
	{\includegraphics[width=6cm]{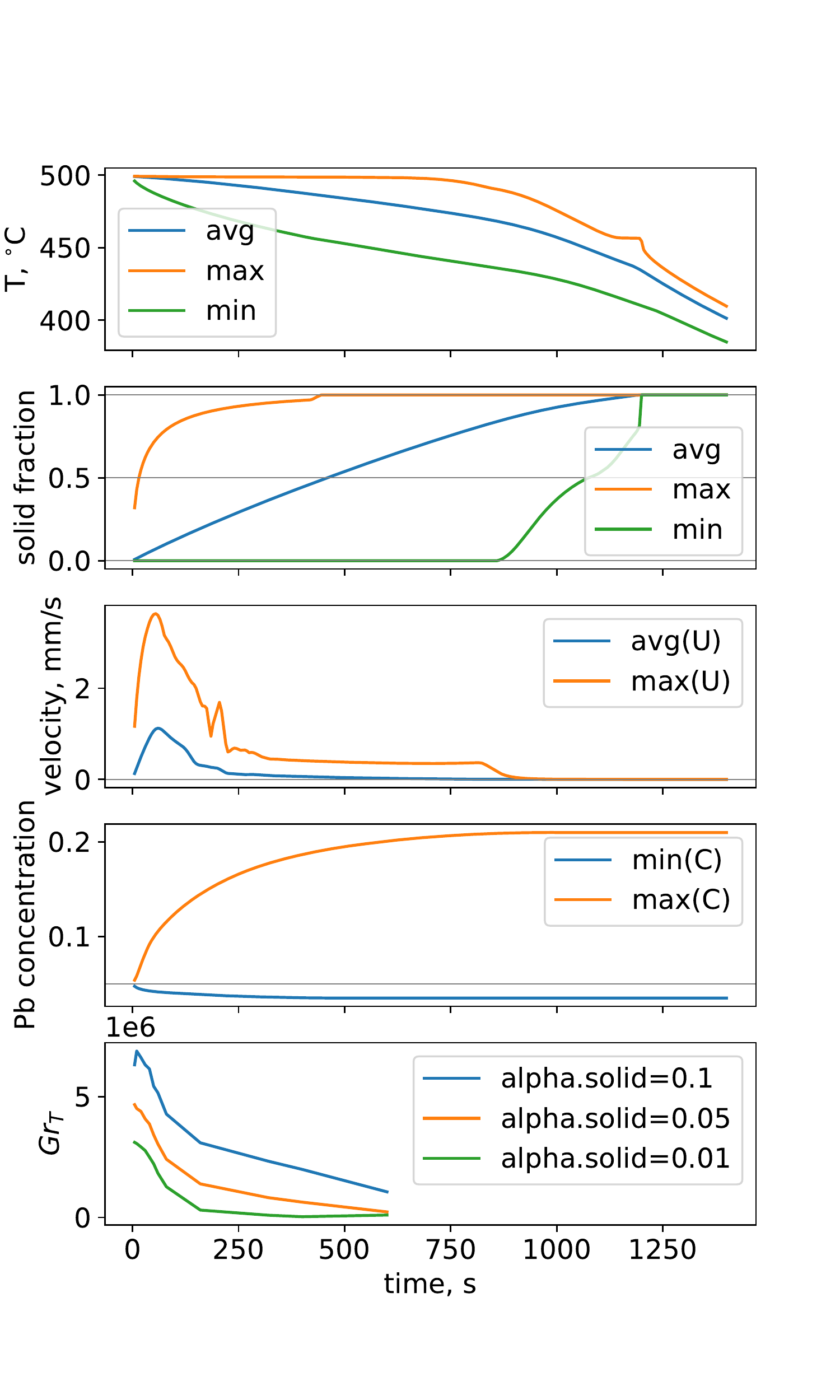}}
	\caption{Dynamics of calculated parameters in Hebditch and Hunt experiment simulation.}
	\label{HH_dynamics}
\end{figure}

Appendix 2: Velocity dynamics.
\begin{figure}[h!]
	\center
	{\includegraphics[width=6cm]{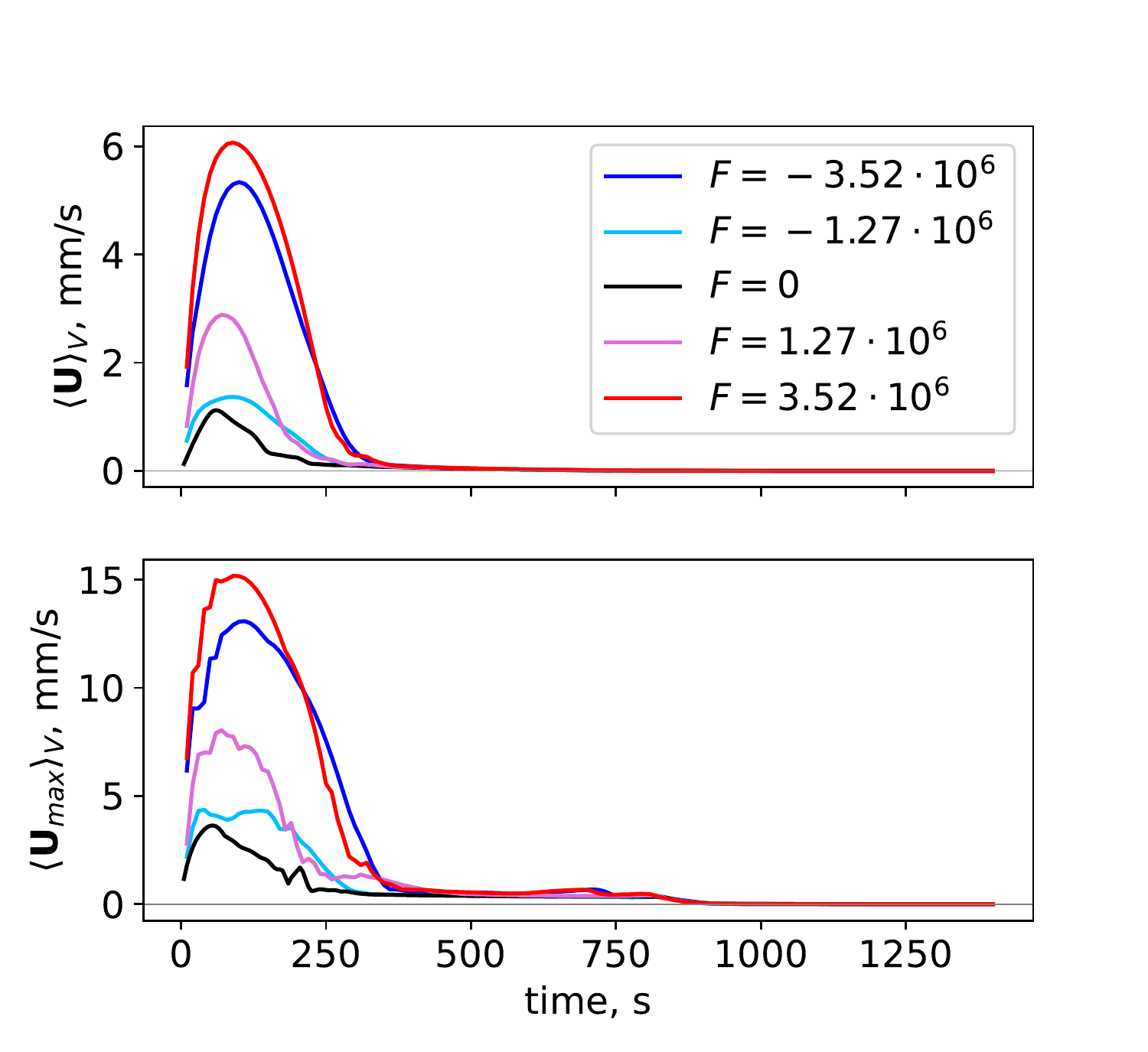}}
	\caption{Dynamics of volume integrated velocity over the solidification time.}
	\label{U(T)}
\end{figure}

Appendix 3: Final concentration graphs.
\begin{figure}[h!]
	\center
	{\includegraphics[width=9cm]{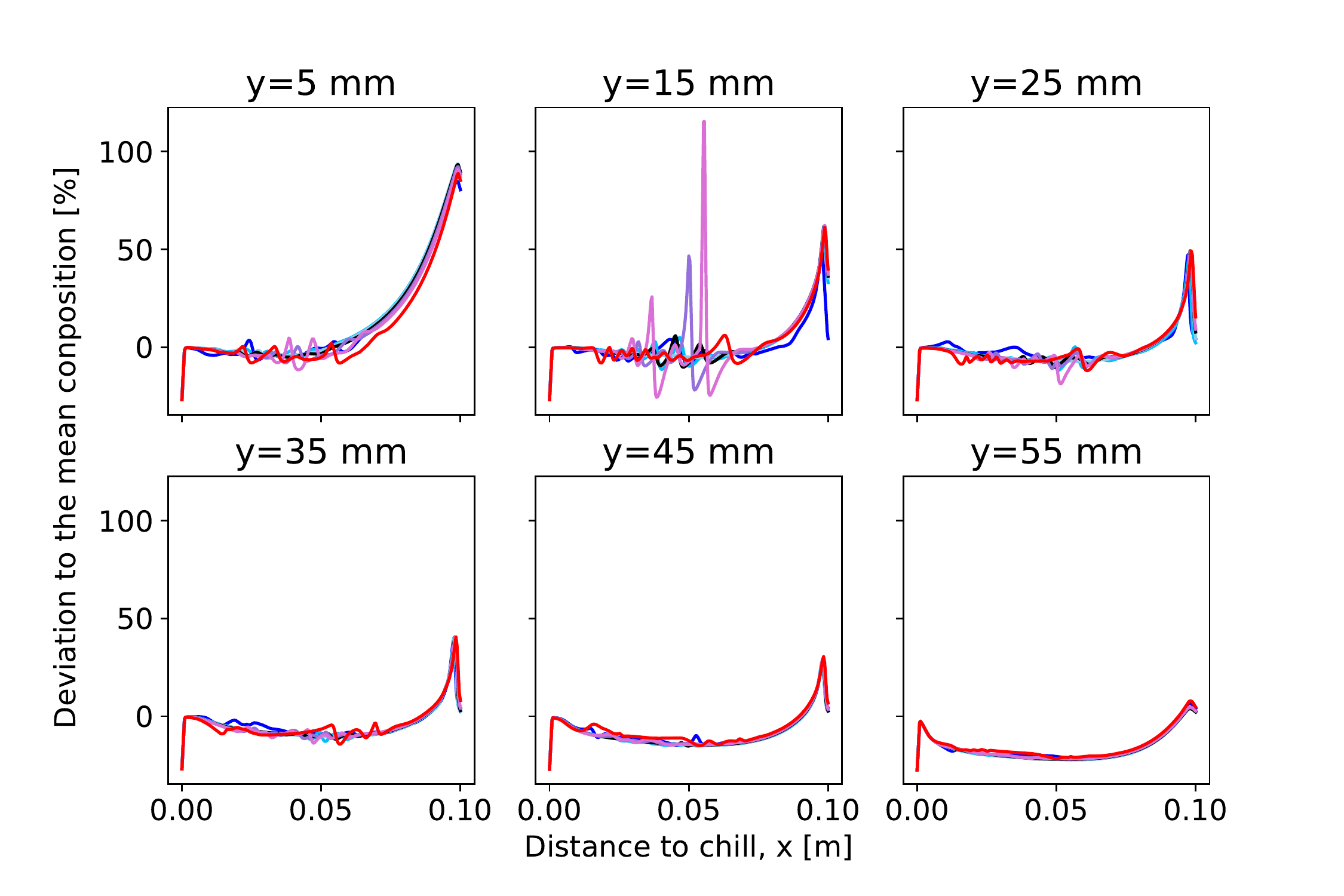}} 
	\includegraphics[width=3cm]{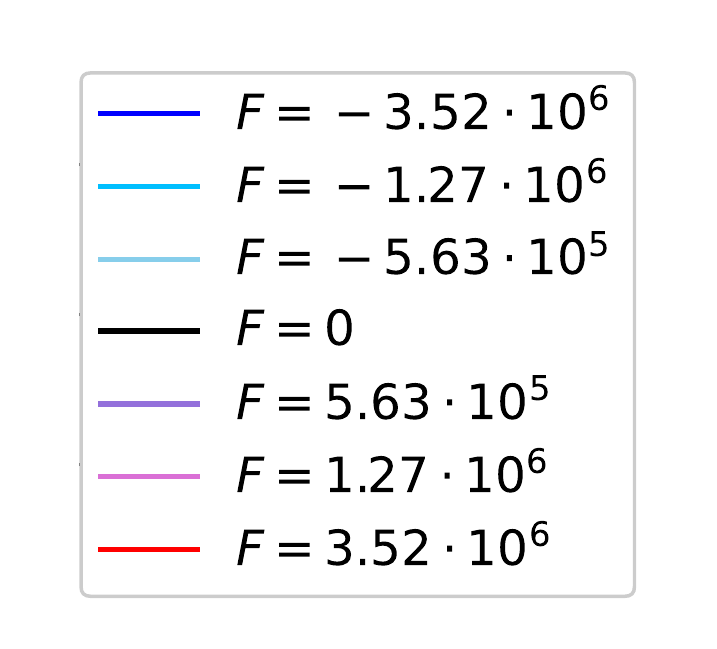}
	\caption{Deviation to the mean composition for different EM-forcing condition.}
	\label{C(x)}
\end{figure}

\end{document}